\documentclass[]{elsarticle}
\makeatletter
\def\ps@pprintTitle{%
	\let\@oddhead\@empty
	\let\@evenhead\@empty
	\def\@oddfoot{}%
	\let\@evenfoot\@oddfoot}
\makeatother
\usepackage{fancyhdr}

\usepackage{epsfig}
\usepackage{amsfonts}
\usepackage{amssymb, graphics, amsmath}
\usepackage{dsfont}
\usepackage{hyperref}
\usepackage{color}
\usepackage{pdflscape}
\usepackage{pifont}
\usepackage{graphicx} 
\usepackage{bm}
\usepackage{slashed}
\usepackage[compat=1.1.0]{tikz-feynman}
\usepackage{tikz}
\usepackage[compat=1.1.0]{tikz-feynhand}

\allowdisplaybreaks

\def\slashchar#1{\setbox0=\hbox{$#1$}           
	\dimen0=\wd0                                    
	\setbox1=\hbox{/} \dimen1=\wd1                  
	\ifdim\dimen0>\dimen1                           
	\rlap{\hbox to \dimen0{\hfil/\hfil}}            
	#1                                             
	\else                                          
	\rlap{\hbox to \dimen1{\hfil$#1$\hfil}}        
	/                                           
	\fi}       

\begin{document}
\title{High-precision determination of the electric and magnetic radius\\ of the proton}

\author[bonn]{Yong-Hui Lin}
\author[darmstadt,emmi]{Hans-Werner Hammer}
\author[bonn,fzj,tbilisi]{Ulf-G. Mei\ss{}ner}

\address[bonn]{Helmholtz-Institut f\"ur Strahlen- und
	Kernphysik and Bethe Center for Theoretical Physics,\\
	Universit\"at Bonn, D-53115 Bonn, Germany}
\address[darmstadt]{Technische Universit\"at Darmstadt, Department of Physics, Institut f\"ur Kernphysik,\\
                    64289 Darmstadt, Germany}
\address[emmi]{ExtreMe Matter Institute EMMI, GSI Helmholtzzentrum f\"ur Schwerionenforschung GmbH,\\
               64291 Darmstadt, Germany}
\address[fzj]{Institute for Advanced Simulation, Institut f\"ur Kernphysik and
	J\"ulich Center for Hadron Physics,\\ Forschungszentrum J\"ulich,
	D-52425 J\"ulich, Germany}
\address[tbilisi]{Tbilisi State University, 0186 Tbilisi, Georgia}

\begin{abstract}
  Using dispersion theory with an improved description of the two-pion continuum
  based on the precise Roy-Steiner analysis of pion-nucleon scattering,
  we analyze recent  data from electron-proton scattering. This allows for
  a high-precision determination of the electric and magnetic radius of the proton,
  $r_E = (0.838^{+0.005}_{-0.004}{}^{+0.004}_{-0.003})\,$fm and
  $r_M = (0.847\pm{0.004}\pm{0.004})\,$fm,
  where the first error refers to the fitting procedure using bootstrap and the data while the second one refers
  to the systematic uncertainty related to the underlying spectral functions. 
\end{abstract}

\maketitle

\thispagestyle{fancy}

\section{Introduction}

The electric radius, $r_E$, and the magnetic radius, $r_M$, of the proton are fundamental
quantities of low-energy QCD, as they are a measure of the probe-dependent size of the proton.
While the electric radius of the proton has attracted much attention in the last decade (see, e.g.,
Refs.~\cite{Carlson:2015jba,Hammer:2019uab,Karr:2020wgh} for recent reviews), this is not
true for its magnetic counterpart, which is not probed in the Lamb shift in electronic or
muonic hydrogen. A major
source of information on the proton form factors (and the corresponding radii) is elastic
electron-proton ($ep$) scattering. These data can be best analyzed in the time-honored
framework of dispersion theory \cite{Chew:1958zjr,Federbush:1958zz,Hohler:1976ax,Mergell:1995bf},
which includes all constraints from unitarity, analyticity
and crossing symmetry and is consistent with the strictures from perturbative QCD at very
large momentum transfer \cite{Lepage:1980fj}. Of particular importance for the proper extraction
of the radii is the
isovector two-pion continuum on the left shoulder of the $\rho$-resonance~\cite{Frazer:1959gy,Hohler:1974eq},
which can be worked out model-independently using dispersively constructed pion-nucleon
scattering amplitudes combined with  data of the pion vector form factor. Based on the
recent Roy-Steiner analysis of pion-nucleon scattering~\cite{Hoferichter:2015hva}, an improved
determination of the two-pion continuum was given in Ref.~\cite{Hoferichter:2016duk}, which also includes 
thorough error estimates. Using a sum rule for the isovector charge radius, in that paper a squared
isovector charge radius, $(r_{E}^v)^2 = 0.405(36)\;$fm$^2$, was obtained 
which is in perfect agreement with a recent state-of-the-art lattice QCD calculation
at physical pion masses, $(r_{E}^v)^2 = 0.400(13)_{\rm sta}(11)_{\rm sys}\;$fm$^2$ \cite{Djukanovic:2021cgp}.
This underscores the importance of the
isovector two-pion continuum for the form factors and demonstrates the consistency of the
new and improved representation from Ref.~\cite{Hoferichter:2016duk} with QCD.
This new representation of the two-pion continuum has so far not been employed in
any dispersion-theoretical analysis of form factor data.

With the advent of new and precise electron-scattering data at
low momentum transfer from Jefferson Laboratory (PRad collaboration)~\cite{Xiong:2019umf},
it is timely to analyze these together with the precise data from the A1 collaboration at
the Mainz Microtron (MAMI)~\cite{Bernauer:2013tpr}
using the improved two-pion continuum contribution. Given the precision of these data
and of the underlying formalism, this will allow for a high-precision determination of
both the electric and the magnetic form factors and the corresponding radii, $r_E$ and
$r_M$, respectively. Clearly, this is an important step in pinning down these fundamental quantities
with high precision. Other recent analyses of the PRad and Mainz data can be found in
Refs.~\cite{Alarcon:2020kcz,Cui:2021vgm,Atac:2021} and will be discussed below.

\section{Formalism}

Here, we briefly summarize the underlying formalism, which is detailed in
Refs.~\cite{Belushkin:2006qa,Lorenz:2014yda}.
The differential cross section for $ep$ scattering can be expressed through the
electric ($G_E$) and magnetic ($G_M $) Sachs form factors (FFs) as
\begin{equation}\label{eq:xs_ros}
\frac{d\sigma}{d\Omega} = \left( \frac{d\sigma}
{d\Omega}\right)_{\rm Mott} \frac{\tau}{\epsilon (1+\tau)}
\left[G_{M}^{2}(t) + \frac{\epsilon}{\tau} G_{E}^{2}(t)\right]\, ,
\end{equation}
where $\epsilon = [1+2(1+\tau)\tan^{2} (\theta/2)]^{-1}$ is the virtual photon
polarization,
$\theta$ is the electron scattering angle in the laboratory frame, $\tau = -t/4m_N^2$,
with $t$ the four-momentum transfer squared and $m_N$ the nucleon mass. Moreover,
$({d\sigma}/{d\Omega})_{\rm Mott}$ is the Mott cross section, which corresponds to scattering off
a point-like spin-1/2 particle. Since $-t\equiv Q^2>0$ is spacelike in $ep$ scattering, the form factors are often
displayed as a function of $Q^2$. Equation~\eqref{eq:xs_ros} will be our basic tool to analyze the data
together with the two-photon corrections from Ref.~\cite{Lorenz:2014yda}.

The electric and magnetic radii of the proton, which are at the center of this investigation,
are given by
\begin{equation}
r_{E/M} = \left(\frac{6}{G_{E/M}(0)}\frac{dG_{E/M}(t)}{dt}\biggr|_{t=0}\right)^{1/2}~.
\end{equation}

For the theoretical analysis, it is advantageous to use the Dirac ($F_1$) and
Pauli ($F_2$) FFs, which are linear combinations of the Sachs FFs:
\begin{equation}
 G_E(t) = F_1(t)-\tau F_2(t)~, ~~~~G_M(t) = F_1(t)+F_2(t)~.
\end{equation}
The FFs for spacelike momentum transfer, $t<0$, are given in terms of an unsubtracted dispersion relation,
\begin{equation}
 F_i(t) = \frac{1}{\pi}\int_{t_0}^{\infty}\frac{\text{Im}F_i(t')dt'}{t'-t}~,\hspace{1cm}i = 1,2~,
\label{disprel}
\end{equation}
with $t_0 = 4M_\pi^2 \, (9M_\pi^2)$ the isovector (isoscalar) threshold, and $M_\pi$ is the
charged pion mass. The spectral functions
are expressed in terms of (effective) vector meson poles and continua, which leads to the following representation
of the FFs:
\begin{align}
F_i^s (t) &= \sum_{V=\omega, \phi ,s_1,s_2,..} \frac{a_i^V}{m_V^2-t} + + F_i^{\pi\rho}(t) + F_i^{\bar{K}K}(t)~,
\nonumber\\
F_i^v (t) &= \sum_{V=v_1,v_2,..}\frac{a_i^V}{m_V^2-t} + F_i^{2\pi}(t)~,
\label{VMDspec}
\end{align}
with $i = 1,2$, in terms of the isoscalar ($s$) and isovector ($v$) components,
$F_i^{(s/v)} = (F_i^p \pm F_i^n)/2$. This representation is advantageous for dispersion
analyses since the intermediate states contributing to the spectral function
have good isospin.
In the isoscalar spectral function, the first two poles correspond to the
$\omega (782)$ and the $\phi (1020)$ mesons, so these masses are fixed and the
residua are bounded as in Ref.~\cite{Lorenz:2014yda}.
Furthermore, we take into account the $\pi\rho$ and
$\bar{K}K$ continua as explained in detail in Ref.~\cite{Belushkin:2006qa}.
The last term in the isovector form factor corresponds to the parameterization
of the two-pion continuum taken from Ref.~\cite{Hoferichter:2015hva}. This is the
essential new theory input compared to earlier dispersive analyses.
The higher mass poles are effective poles that parameterize the spectral function
at large $t$. We explicitly check in our analysis that the radii are
insensitive to the details of this parameterization. The fit parameters are therefore
the various vector meson residua $a_i^V$ and the masses of the additional vector
mesons $s_i, v_i$.  Note that from the proton data alone, the isospin of a given pole
  is not determined. We simply assign a given number of isoscalar and isovector poles
  besides the continnum contributions, which have a given isospin, as well as the $\omega$
  and $\phi$ mesons (see also the discussion in the Appendix).
In addition, we  fulfill the normalization conditions $F_1(0)= 1$
(in units of the elementary charge $e$)
and $F_2(0) = \mu_p$, with $\mu_p$ the anomalous magnetic moment of the proton.
To ensure the stability of the fit \cite{SabbaStefanescu:1978hvt},
we demand that the residua of the vector meson poles are bounded, $|a_i^V| < 5\,$GeV$^2$, and
that no effective poles with masses below 1 GeV appear.
Finally, the FFs must satisfy the superconvergence relations
\begin{equation}
\int_{t_0}^\infty {\rm Im} F_i(t)t^n dt = 0~, ~~i = 1,2~,
\end{equation}  
with $n=0$ for $F_1$ and $n=0,1$ for $F_2$, corresponding to the fall-off with inverse
powers of $Q^2$ at large momentum transfer as demanded by perturbative QCD \cite{Lepage:1980fj}.

These parameterizations \eqref{VMDspec} are used in Eq.~\eqref{eq:xs_ros} and the number
of isoscalar and isovector poles is determined by the condition to obtain the best fit
to the data. The quality of the fits is measured in terms of the traditional $\chi^2$,
\begin{equation}
\chi^2_1 = \sum_i\sum_k\frac{(n_k C_i - C(Q^2_i,\theta_i,\vec{p}\,))^2}{(\sigma_i+\nu_i)^2}~,
\label{chi1}
\end{equation}
where $C_i$ are the cross section data at the points $Q^2_i,\theta_i$ and $C(Q^2_i,\theta_i,\vec{p}\,)$
are the cross sections for a given FF parameterization for the parameter values contained in $\vec{p}$.
Moreover, $n_k$ are normalization coefficients for the various data sets (labelled by the integer $k$),
while  $\sigma_i$ and $\nu_i$ are their statistical and systematical errors, respectively. A more refined
definition of the $\chi^2$ is given by~\cite{Lorenz:2014yda}
\begin{align}
\chi^2_2 = \sum_{i,j}\sum_k(n_k C_i - C(Q^2_i,\theta_i,\vec{p}\,))[V^{-1}]_{ij}(n_k C_j
- C(Q^2_j,\theta_j,\vec{p}\,))~,
\label{chi2}
\end{align}
in terms of the covariance matrix $V_{ij} = \sigma_i\sigma_j\delta_{ij} + \nu_i\nu_j$.
Theoretical errors will be calculated on the one hand using a bootstrap method.
We simulate a large number of data sets by randomly varying the points in the
original set within the given errors assuming their normal distribution. We then fit to each of them separately,
derive the radius from each fit, and analyze the distribution of these radius values
(see App.~D of Ref.~\cite{Lorenz:2014yda} for details).
On the other hand theoretical errors are estimated by varying the number of effective vector meson poles.
The first error thus gives
the uncertainty due to the fitting procedure (bootstrap) and the data while the second one reflects the accuracy of the
spectral functions underlying the dispersion-theoretical analysis. Note that these two errors are not in a strict one-to-one
correspondence to the commonly given statistical and systematic errors.
We further remark that since the effect of the two-photon corrections on the extracted radii is
minimal, we do not attempt to quantify their uncertainty here (see also
the discussion in~\cite{Lorenz:2012tm} and references therein).
  
\section{Results}

As a first validation of our method, we only consider the PRad data~\cite{Xiong:2019umf}.
These can be best fitted with the lowest two isoscalar mesons (the $\omega$ and the $\phi$)
and two additional isovector ones
($2s + 2v$ poles). Fitting with statistical errors only (as in Ref.~\cite{Xiong:2019umf}), we have
a  $\chi^2/{\rm dof} = 1.33$, completely consistent with the results reported there. Including
also the systematic errors, the reduced $\chi^2$ is slightly improved and we find as central values
\begin{equation}
\label{eq:rPRad}  
r_E = (0.829\pm 0.012\pm 0.001)\,{\rm fm}~, ~~ 
r_M = (0.843\pm 0.007^{+0.018}_{-0.012})\,{\rm fm}~,
\end{equation}
consistent with the PRad result,
$r_E = (0.831\pm 0.007_{\rm stat}\pm 0.012_{\rm syst})\,$fm for the electric radius. In our case, the first
error is obtained by  bootstrap using 1000 samples and the second error is obtained by varying
the number of poles from two isoscalar and two isovector ones (which gives the best solution)
up to 5 isoscalar plus 5 isovector poles. While the absolute $\chi^2$ of these 8 different
solutions is almost the same, the $\chi^2/{\rm dof}$ increase from $1.33$ to $1.61$. Note also that
the uncertainty in the magnetic radius is sizeably larger than the one of the electric radius,
which is due to the fact that at the very low $Q^2$ probed by PRad, the electric form factor
dominates the cross section. We remark that the PRad data have also been analyzed in Ref.~\cite{Paz:2020prs}
using the $z$-expansion, which also finds a sizeably  increased statistical error as done here.

\begin{figure}[tb!]
	\centering
	\includegraphics[width=0.675\textwidth]{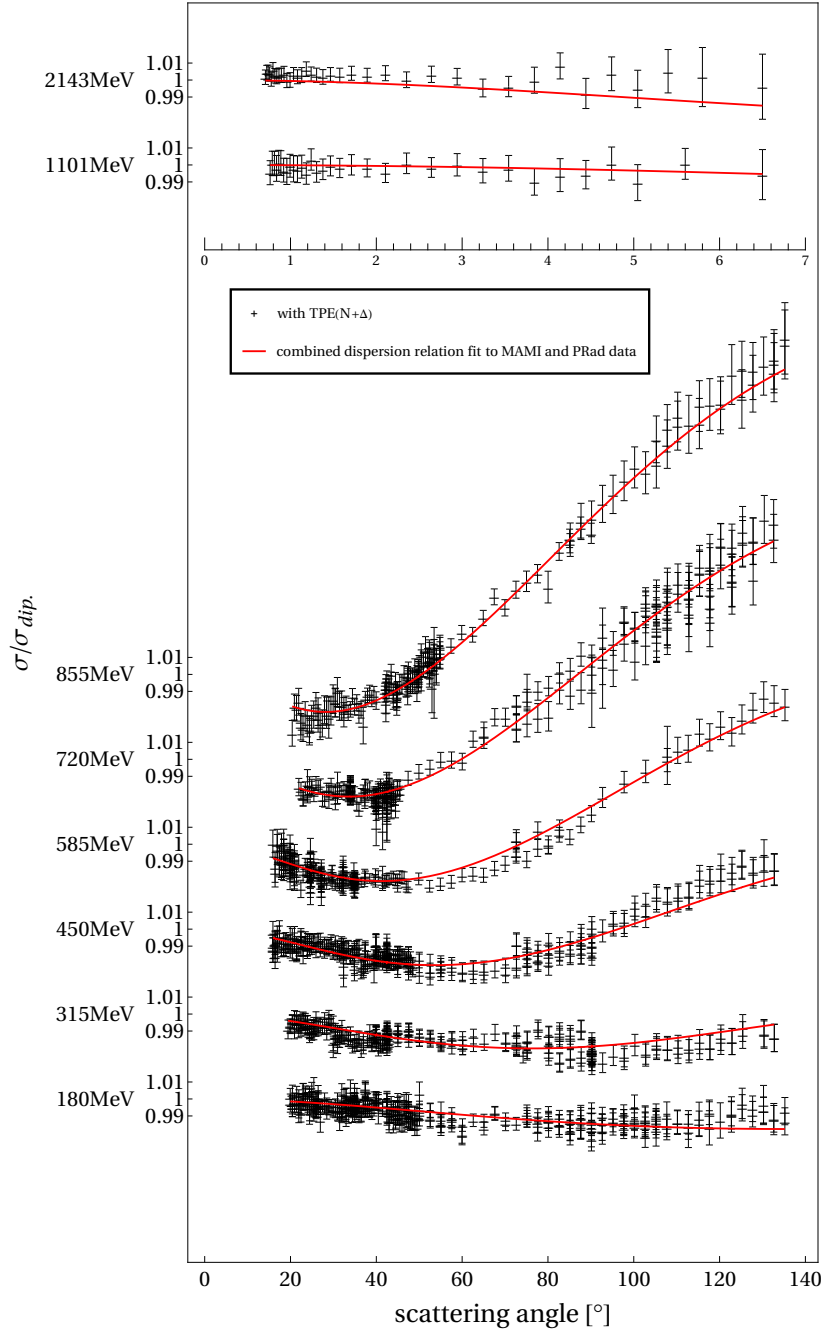}
	\caption{Combined fit to the PRad (upper panel) and the Mainz data (lower panel)
          as described in the text. The dipole cross section $\sigma_{\rm dip.}$ is obtained
          by using the dipole approximation to the FFs.
         \label{fig:fits}}
\end{figure}

Next, we turn to the combined analysis of the Mainz and the PRad data. We note that we increase
the weight of the PRad data by a factor ten in the combined $\chi^2$. This is legitimate as the
PRad data probe much smaller momentum transfer than the Mainz data and thus should be enhanced.
Changing this weight by a factor of two leads to changes in the proton radii that are well
covered by the uncertainties discussed below. The best fit to these
data is shown in Fig.~\ref{fig:fits} for the PRad data and the Mainz data (normalized to the
dipole cross section $\sigma_{\rm dip}$).
To best describe these combined data sets requires $5s+5v$ poles (as in Ref.~\cite{Lorenz:2014yda}
for the MAMI data alone)  with a $\chi^2/{\rm dof} = 1.25$. The corresponding vector meson
parameters (masses, residua) and the normalization constants of the various data sets are
collected and discussed in~\ref{app:A}. In such a combined fit, the PRad data are described slightly worse than
before, as the Mainz data set is much larger and has also smaller error bars.
The resulting radii are:
\begin{eqnarray}
\label{eq:fullfit}  
r_E &=& (0.838^{+0.005}_{-0.004}{}^{+0.004}_{-0.003})\,{\rm fm}~, ~~\nonumber \\
r_M &=& (0.847\pm{0.004}\pm{0.004})\,{\rm fm}~.
\end{eqnarray}
These results are consistent with our earlier dispersion-theoretical determinations
\cite{Lorenz:2012tm,Lorenz:2014yda} but have much improved and smaller uncertainties.
In particular, the fit error based on bootstrap (using 5000 samples)
is considerably improved as compared to
Ref.~\cite{Lorenz:2014yda} which is related to the new PRad data, while the fit error
for the magnetic radius is similar (as it is dominated by the Mainz data). The theory
error estimate is obtained performing 11 different sets of fits with varying numbers
of isoscalar and isovector poles, from $2s+2v$ to $7s+7v$ poles, where the reduced $\chi^2$/dof
only varies by less than 3\% from the optimal value obtained for $5s+5v$ poles.
Note further that the theory uncertainty
is improved compared to the one in Ref.~\cite{Lorenz:2012tm}, with the differences
that we do not include neutron data here and a much better and precise determination
of the so important isovector two-pion continuum is employed.

In Table~\ref{tab:radii} we have collected these results together with other
recent determinations of $r_E$ and $r_M$ employing both the PRad and the Mainz data.
Within the quoted errors all extracted radii are consistent.
The  analysis in Ref.~\cite{Alarcon:2020kcz} is similar to ours. In contrast to our approach,
however, it employs a dispersively improved chiral perturbation theory representation of
the two-pion continuum. This  approach is subject to uncertainties in the $\rho$-region as stressed
in Ref.~\cite{Leupold:2017ngs}, different from the exact representation used here.
The work of Ref.~~\cite{Cui:2021vgm}
can not be directly compared, as it is based on a continued fraction approach which has no
relation to the dispersion-theoretical method used here. Also, in that paper no value
for the magnetic radius is given. Similar remarks hold for the work of Ref.~\cite{Atac:2021},
which applies various fit functions (not guided by unitarity) to the flavor-dependent Dirac
form factors with $Q^2 \leq 1$~GeV$^2$ to extract the proton and the neutron charge radii.
We note that our charge radius is also consistent with the current CODATA
value, $r_E=0.8414(19)$\,fm \cite{CODATAnew}. Moreover, it is in agreement with the
value $r_E=0.827(20)$ fm obtained by combining the recent precise lattice result by
Djukanovic et al. for the isovector radius with the experimental neutron charge radius
\cite{Djukanovic:2021cgp,Filin:2019eoe}. Another recent
lattice study calculated the proton charge radius directly but neglected
disconnected contributions to the isoscalar current, which are computationally very
expensive \cite{Alexandrou:2020aja}. Because of this, the charge radius comes
out smaller but their isovector result is in agreement with Ref.~\cite{Djukanovic:2021cgp} and the
experimental value.\footnote{For a detailed discussion of previous lattice
QCD calculations we refer to Refs.~\cite{Djukanovic:2021cgp,Alexandrou:2020aja}.}
Thus finally a consistent picture for the proton charge radius appears to
emerge \cite{Hammer:2019uab}.

\begin{table}
\begin{center}
\begin{tabular}{|c|c|c|c|c|}
\hline
           & This work & Ref.~\cite{Alarcon:2020kcz} & Ref.~\cite{Cui:2021vgm} & Ref.~\cite{Atac:2021} \\
\hline
$r_E$ [fm] & $0.838^{+0.005}_{-0.004}{}^{+0.004}_{-0.003}$ & $0.842\pm 0.002_{\rm fit}\pm 0.010_{\rm th}$
                                  & $0.847 \pm0.008_{\rm sta}$ & $0.852 \pm0.002_{\rm sta}\pm 0.009_{\rm sys}$\\
$r_M$ [fm] & $0.847\pm{0.004}\pm{0.004}$ & $0.850\pm 0.001_{\rm fit}\pm 0.010_{\rm th}$
                                                         & --- & ---\\
\hline
\end{tabular}
\caption{Comparison of the proton radii extracted in this work and other recent papers.
  For the definitions of the errors in Refs.~\cite{Alarcon:2020kcz,Cui:2021vgm,Atac:2021}, see these
  papers.}
\label{tab:radii}
\end{center}
\end{table}

\begin{figure}[t!]
	\centering
	\includegraphics[width=0.45\textwidth]{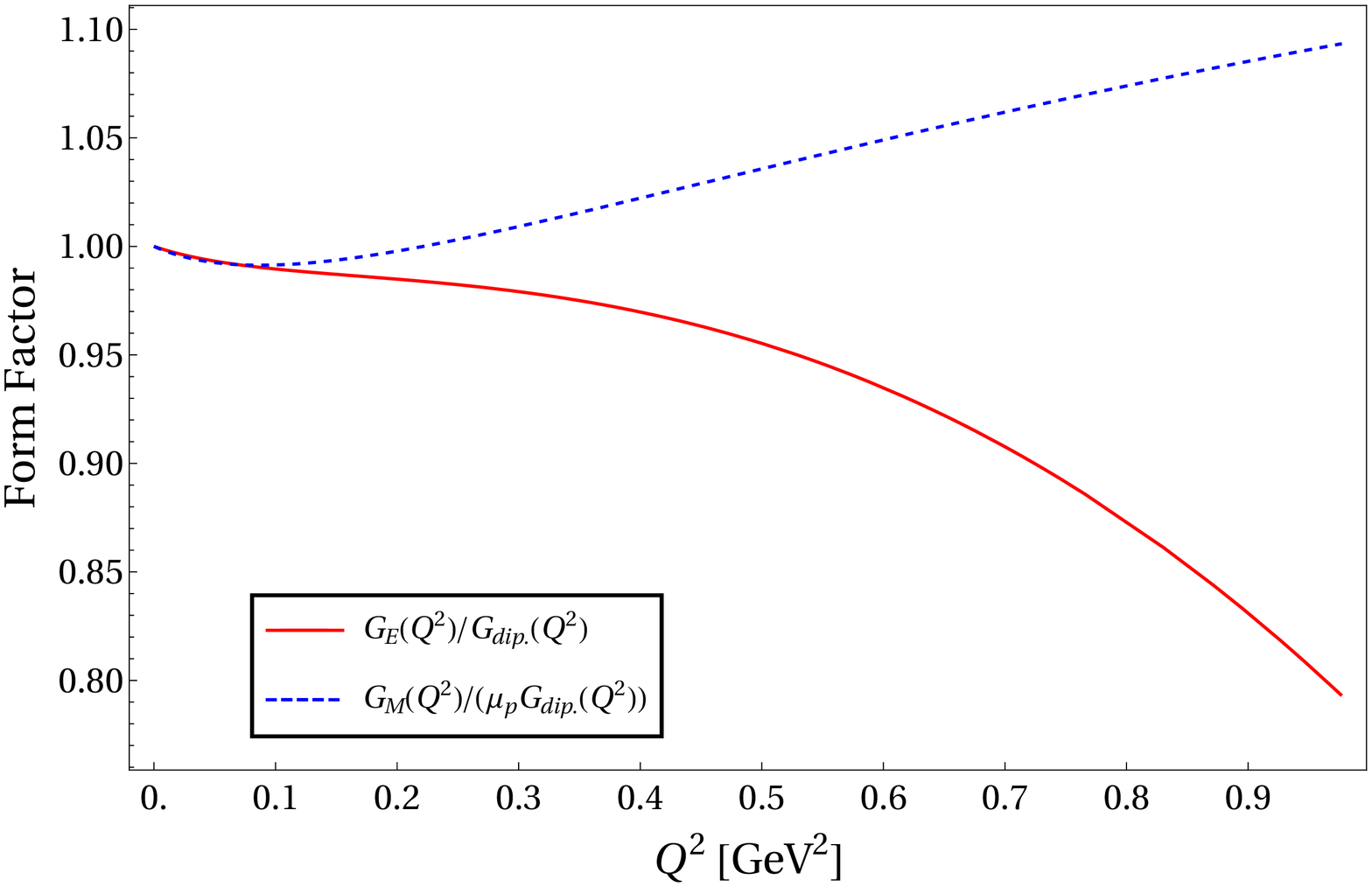}~~~~~
	\includegraphics[width=0.45\textwidth]{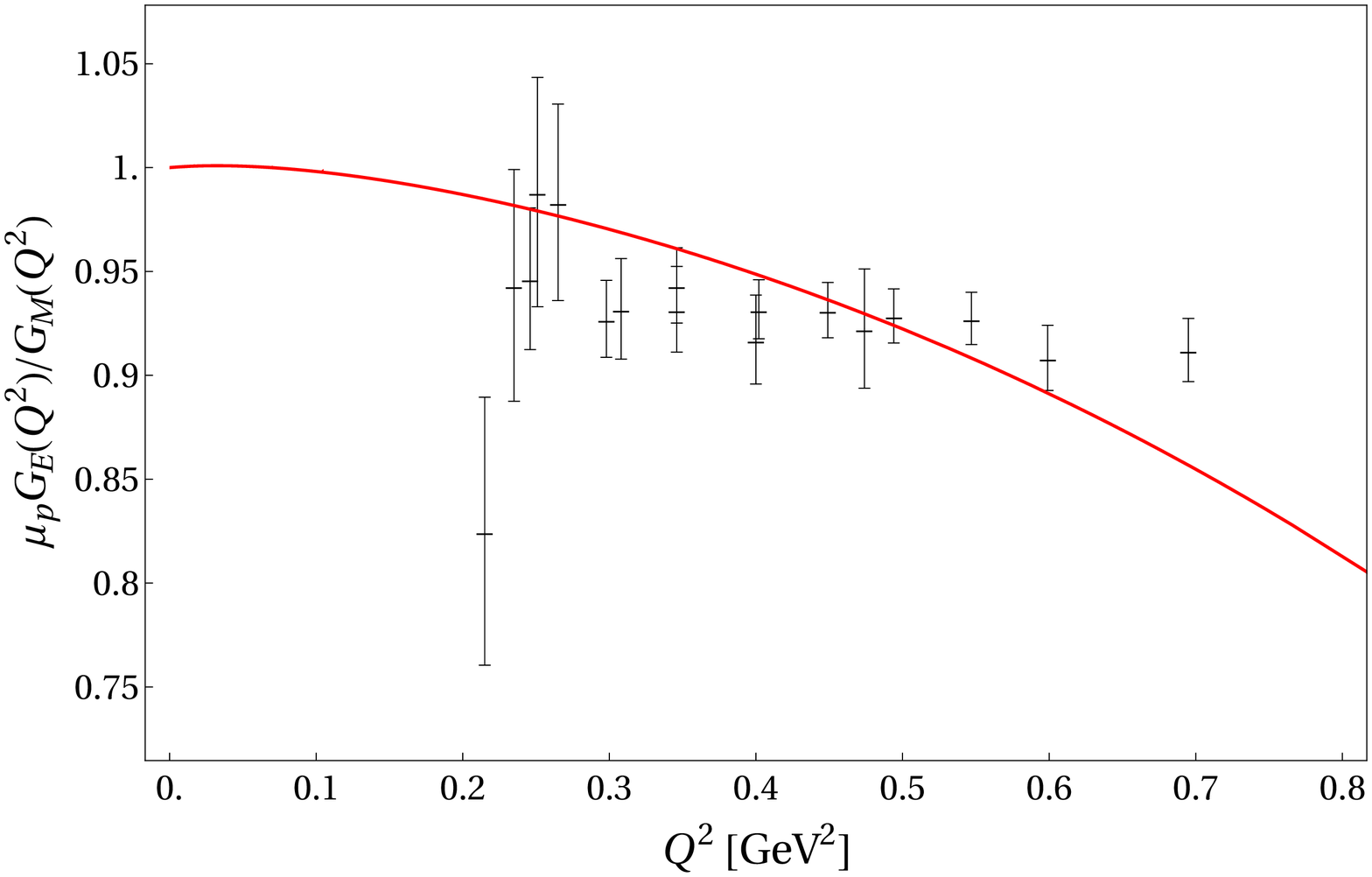}
	\caption{Left panel: Electric (red solid line) and magnetic (blue dashed line) FF
          normalized to the dipole FF for our best solution, $G_E(Q^2)/G_{\rm dip}(Q^2)$ and
          $G_M(Q^2)/(\mu_P G_{\rm dip}(Q^2)$, respectively. Right panel: FF ratio $\mu_p G_E(Q^2)/G_M(Q^2)$
          (red solid line) compared to the data of Refs.~\cite{Ron:2011rd,Zhan:2011ji}.
           \label{fig:GEM}}
\end{figure}

In Fig.~\ref{fig:GEM} (left panel), we show the resulting electric and magnetic FF of the proton normalized
to the dipole FF, $G_{\rm dip}(Q^2) = (1+Q^2/Q^2_{\rm dip})^{-2}$, with $Q^2_{\rm dip} = 0.71\,$GeV$^2$.
These show a behavior similar to what was
found in earlier dispersion-theoretical studies, and we note that again physically constrained
fits do not produce oscillations as seen, e.g., the the magnetic form factor in Ref.~\cite{Bernauer:2013tpr}.
We refrain here from displaying the corresponding uncertainties estimates to avoid clutter.
In the right panel of Fig.~\ref{fig:GEM}, the FF ratio $\mu_p G_E(Q^2)/G_M(Q^2)$ is displayed together
with the data of Refs.~\cite{Ron:2011rd,Zhan:2011ji}. Our ratio is consistent with these data, which
were not included in the fits.

\section{Summary}

In this paper, we have continued the dispersion-theoretical analysis of the proton form factors
triggered by two main developments. On the theoretical side, a much improved representation of
the two-pion-continuum contribution to the isovector spectral function based on the precision
results from the Roy-Steiner analysis of pion-nucleon scattering has been presented~\cite{Hoferichter:2015hva}. On the experimental side, new $ep$ scattering data at very
low $Q^2$ from the PRad collaboration~\cite{Xiong:2019umf} have
become available. Using our improved spectral functions and employing the two-photon
corrections worked out in Ref.~\cite{Lorenz:2014yda}, we have analyzed these new data
as well as the combination of the PRad and the Mainz data~\cite{Bernauer:2013tpr}
which allowed us to extract the proton's electric and magnetic radius with unprecedented precision,
as given in Eq.~\eqref{eq:fullfit}. Theoretical uncertainties from the fit procedure and from
variations in the spectral functions have been worked out. In the future, Bayesian methods will
be used to further improve these uncertainty estimates. Furthermore, fits including also the
time-like proton form factor data should be performed. Finally, data
for the scattering off the neutron should be included, as these can be used to precisely pin
down the corresponding neutron radii using chiral effective field theory for few-nucleon
systems~\cite{Filin:2019eoe}.

\section*{Acknowledgements}

This work of UGM and YHL is supported in
part by the NSFC and the Deutsche Forschungsgemeinschaft (DFG, German Research
Foundation) through the funds provided to the Sino-German Collaborative  
Research Center TRR110 ``Symmetries and the Emergence of Structure in QCD''
(NSFC Grant No. 12070131001, DFG Project-ID 196253076 - TRR 110),
by the Chinese Academy of Sciences (CAS) through a President's
International Fellowship Initiative (PIFI) (Grant No. 2018DM0034), by the VolkswagenStiftung
(Grant No. 93562), and by the EU Horizon 2020 research and innovation programme, STRONG-2020 project
under grant agreement No. 824093. HWH was supported by the Deutsche Forschungsgemeinschaft (DFG, German
Research Foundation) -- Projektnummer 279384907 -- CRC 1245
and by the German Federal Ministry of Education and Research (BMBF) (Grant
No. 05P18RDFN1).

\pagebreak

\appendix
\section{Fit parameters and discussion}
\label{app:A}

We collect the various vector meson masses and couplings that appear in the
spectral functions Eqs.~\eqref{VMDspec} and the normalization constants of the various
data sets (see Ref.~\cite{Lorenz:2014yda} for precise definitions) in Table~\ref{tab:values}.
\begin{table}[ht!]
	\centering
	\begin{tabular}{|l|c|c|c||l|c|c|c|}
		\hline
		$V_{s}$ & $m_V$ & $a_1^V$ &$a_2^V$ & $V_{v}$ & $m_V$ & $a_1^V$ & $a_2^V$  \\
		\hline
		$\omega$ & $0.7830$	& $0.8572$	& $0.0177$	& $v_1$ & $1.0426$	 & $0.6876$ & $-1.5086$	\\
		$\phi$   & $1.0190$	& $-1.3155$	& $0.9955$	& $v_2$ & $2.3839$	 & $-4.3848$ & $4.8535$	\\
		$s_1$    & $1.4790$	& $ 2.6928$	& $-4.8054$	& $v_3$ & $3.3482$	 & $-3.6869$ & $-4.6070$	\\
		$s_2$    & $2.2381$	& $1.6155$	& $4.8620$	& $v_4$ & $3.5665$	 & $2.4907$ & $-3.0026$	\\
		$s_3$    & $3.4614$	& $-2.9454$	& $-1.2582$	& $v_5$ & $4.7887$	 & $4.7612$ & $2.7502$	\\
		\hline
	\end{tabular}
\begin{tabular}{|l|l|l|l|l|l|l|l|l|l|l|l|l|l|l|}
\hline
n1 & $0.9965$ & n6 & $0.9909$ & n11 & $1.0000$ & n16 & $1.0019$ & n21 & $0.9999$ & n26 & $1.0041$
   & n31 & $0.9980$  \\
n2 & $1.0066$ & n7 & $0.9983$ & n12 & $1.0036$ & n17 & $1.0013$ & n22 & $0.9900$ & n27 & $1.0100$
   & $\tilde{\rm n}1$ & $0.9989$  \\
n3 & $1.0028$ & n8 & $0.9937$ & n13 & $1.0039$ & n18 & $1.0026$ & n23 & $1.0033$ & n28 & $1.0100$
   & $\tilde{\rm n}2$ & $1.0056$  \\
n4 & $1.0011$ & n9  & $1.0080$ & n14 & $1.0057$ & n19 & $1.0014$ & n24 & $1.0075$ & n29 & $0.9992$& &\\
n5 & $1.0037$ & n10 & $1.0000$ & n15 & $1.0065$ & n20 & $1.0053$ & n25 & $1.0088$ & n30 & $1.0069$& & \\
\hline
\end{tabular}
\caption{The parameters obtained from the fit to the combined PRad and  MAMI data based on dispersion
  relations: Vector meson (upper panel) and normalization (lower panel) parameters. The normalization
  constants n1$,\ldots ,$n31 refer to the MAMI data sets, whereas $\mathrm{\tilde{\rm n}1,\tilde{\rm n}2}$
  normalize the PRad data. Masses $m_V$ are given in GeV and couplings $a_i^V$ in GeV$^{2}$.}
	\label{tab:values}
\end{table}

Some remarks on the results presented in this table are in order. As in earlier dispersion-theoertical
analyses, see e.g. \cite{Hohler:1976ax,Mergell:1995bf,Belushkin:2006qa}, we find no OZI suppression for the
couplings of the $\phi$. However, there is some close-by pole (here, $v_1$) which cancels a large part
of the $\phi$ contribution (as noted in the main text, one cannot perform an isospin separation fitting
only proton data). This will certainly change once neutron data are included, see e.g. \cite{Belushkin:2006qa}.
For that reason, we also did not consider the uncertainties of the $\bar{K}K$ and $\pi\rho$ continua, as
there are a) sizeable cancellations in this mass region  and b) this region is of minor importance for the
radius extraction.  We also note that the tensor coupling of the $\omega$ is expected to be suppressed in
vector meson dominance  due to the smallness of the isoscalar
nucleon anomalous magnetic moment. This we indeed confirm consistently with the earlier works
\cite{Hohler:1976ax,Mergell:1995bf,Belushkin:2006qa,Lorenz:2012tm,Lorenz:2014yda}.

\end{document}